\documentclass{emulateapj}
\usepackage{natbib}
\usepackage{epsfig}
\usepackage{graphicx}
\usepackage{subfigure}
\usepackage{float}
\usepackage{amsmath}
\usepackage{color}
\usepackage{amssymb}
\usepackage{amsfonts}
\usepackage{apjfonts}
\usepackage[colorlinks,linkcolor=blue,anchorcolor=green,citecolor=blue]{hyperref}
\shorttitle{GRB 160625B As a New Probe of LIV}
\shortauthors{Wei et al.}
\begin{document}

\title{A New Test of Lorentz Invariance Violation: the Spectral Lag Transition of GRB 160625B}

\author{Jun-Jie Wei\altaffilmark{1,2}, Bin-Bin Zhang\altaffilmark{3,4}, Lang Shao\altaffilmark{5,1}, Xue-Feng Wu\altaffilmark{1,6,7}, and Peter M{\'e}sz{\'a}ros\altaffilmark{8,9,10}}

\affil{$^1$Purple Mountain Observatory, Chinese Academy of Sciences, Nanjing 210008, China; xfwu@pmo.ac.cn \\
$^2$ Guangxi Key Laboratory for Relativistic Astrophysics, Nanning 530004, China \\
$^3$Instituto de Astrof\'isica de Andaluc\'a (IAA-CSIC), P.O. Box 03004, E-18080 Granada, Spain \\
$^4$Scientist Support LLC, Madsion, AL 35758, USA \\
$^5$Department of Space Sciences and Astronomy, Hebei Normal University, Shijiazhuang 050024, China\\
$^6$School of Astronomy and Space Science, University of Science and Technology of China, Hefei, Anhui 230026, China\\
$^7$Joint Center for Particle, Nuclear Physics and Cosmology, Nanjing University-Purple Mountain Observatory, Nanjing 210008, China\\
$^8$Department of Astronomy and Astrophysics, Pennsylvania State University, 525 Davey Laboratory, University Park, PA 16802\\
$^9$Department of Physics, Pennsylvania State University, 104 Davey Laboratory, University Park, PA 16802\\
$^{10}$Center for Particle and Gravitational Astrophysics, Institute for Gravitation and the Cosmos, Pennsylvania State University, 525 Davey
Laboratory, University Park, PA 16802}


\begin{abstract}
Possible violations of Lorentz invariance (LIV) have been investigated for a long time using the observed spectral
lags of gamma-ray bursts (GRBs). However, these generally have relied on using a single photon in the highest energy
range. Furthermore, the search for LIV lags has been hindered by our ignorance concerning the intrinsic time lag
in different energy bands. GRB 160625B, the only burst so far with a well-defined transition from $positive$
lags to $negative$ lags provides a unique opportunity to put new constraints on LIV. Using multi-photon
energy bands we consider the contributions to the observed spectral lag from both the intrinsic time lag and
the lag by LIV effects, and assuming the intrinsic time lag to have a positive dependence on the photon energy,
we obtain robust limits on LIV by directly fitting the spectral lag data of GRB 160625B. Here we show that
these robust limits on the quantum gravity energy scales are $E_{\rm QG,1}\geq0.5\times10^{16}$ GeV for the linear,
and $E_{\rm QG,2}\geq1.4\times10^{7}$ GeV for the quadratic LIV effects, respectively. In addition, we give for
the first time a reasonable formulation of the intrinsic energy-dependent time lag.
\end{abstract}

\keywords{astroparticle physics --- gamma-ray burst: individual (GRB 160625B) --- gravitation}

\section{Introduction}
\label{sec:intro}
Spectral lag, the arrival time delay between light curves in different energy bands (or
between correlated photons with different energies), is a common feature in gamma-ray bursts
(GRBs; e.g., \citealt{1995A&A...300..746C,1996ApJ...459..393N,1997ApJ...486..928B}).
Generally speaking, most GRBs show \emph{positive} lags, i.e., light curves at higher energies
(say, in the MeV range) peak earlier than those at lower energies (say, in a range of 10 s to 100 s keV).
However, there some rare cases showing zero lags or even \emph{negative} lags (e.g., \citealt{2000ApJ...534..248N,
2006ApJ...646..351L,2012MNRAS.419..614U}). Since the launch of the \emph{Fermi} satellite,
many GRBs with high energy emission above 100 MeV have been detected. In contrast to the \emph{positive}
lags of low energy emission, GeV photons are found delayed with respect to MeV photons in many
(but not all) GRBs (i.e., \emph{negative} lags; see \citealt{2009Sci...323.1688A,2009Natur.462..331A,2011ApJ...733L..40M}).
Some physical models have been formulated to account for the intrinsic origin of lags (e.g.,
\citealt{2001ApJ...554L.163I,2005MNRAS.362...59S,2009ApJ...707.1404T,2016ApJ...825...97U}).
Particularly, \cite{2016ApJ...825...97U} showed that the intrinsic \emph{positive} lags could be well
reproduced by a simple model invoking synchrotron radiation from a rapidly expanding outflow.

On the other hand, one possible explanation for the \emph{negative} lags is provided by Quantum Gravity (QG)
effects. One such effect is the Lorentz invariance violation (LIV). Lorentz invariance is typically expected
to be broken at the Plank scale ($E_{\rm QG}\approx E_{\rm Pl}=\sqrt{\hbar c^{5}/G}\simeq1.22\times10^{19}$ GeV;
see \citealt{2005LRR.....8....5M,2013LRR....16....5A}, and references therein).\footnote{Note that
here we adopted a LIV scenario with broken relativistic symmetries. This reflects the earlier incarnations of the
relevant phenomenology on the QG side, although in recent years more attention has focused on a
description of these QG effects in a Doubly Special Relativity (DSR) scenario, in which
relativistic symmetries are deformed rather than broken (e.g.,
\citealt{2002IJMPD..11...35A,2002Natur.418...34A,2002PhLB..539..126K,2003PhRvD..67d4017M}).}
Many theories of QG predict that LIV happens at high energy scales, since high energy photons may interact with
the foamy structure of space-time at small spatial scales \citep{1997IJMPA..12..607A}. In such cases, the speed
of light in vacuum would depend on the energy of the photon, and high energy photons propagate
in vacuum slower than low energy photons \citep{1998Natur.393..763A}. The energy scale for LIV, $E_{\rm QG}$,
could therefore be constrained by the arrival time differences of the photons with different energies
originating from the same astronomical source \citep{1998Natur.393..763A,2013APh....43...50E}.

Thanks to their short spectral lags, cosmological distances, and very high energetic photons,
GRBs have been viewed as the most promising sources for studying the LIV effects
\citep{1998Natur.393..763A,2006APh....25..402E,2008JCAP...01..031J}. In the past, various limits on LIV have
been obtained from the spectral time lags of individual GRB or a large sample of GRBs (e.g., \citealt{1998Natur.393..763A,
1999PhRvD..59k6008C,1999PhRvL..82.4964S,2003A&A...402..409E,2006APh....25..402E,2004ApJ...611L..77B,
2006PhLB..643...81K,2008JCAP...01..031J,2009Sci...323.1688A,2009Natur.462..331A,2009CQGra..26l5007B,
2009PhRvD..80k6005X,2010APh....33..312S,2012APh....36...47C,2016ChPhC..40d5102C,2012PhRvL.108w1103N,
2013APh....43...50E,2013PhRvL.110t1601K,2013PhRvD..87l2001V,2015NatPh..11..344V,2015ApJ...808...78P,
2015APh....61..108Z,2016JCAP...08..031W}).
In particular, \cite{2009Sci...323.1688A} used the time lag of the highest energy (13.2 GeV) photon from
GRB 080916C to constrain the linear LIV energy scale ($E_{\rm QG,1}$) and presented a stringent limit of
$1.3\times10^{18}$ GeV, improving the previous limits by at least one order of magnitude.
\cite{2009Natur.462..331A} set the current strictest limits on both the linear and quadratic LIV
energy scales by analyzing the arrival time delay between a 31 GeV photon and the low energy (trigger)
photons from GRB 090510. The limits set are $E_{\rm QG,1}>(1-10)E_{\rm Pl}$
and $E_{\rm QG,2}>1.3\times10^{11}$ GeV. However, these limits were based on the rough time lag of a
single GeV-scale photon. It is necessary to consider using the true spectral time lags of bunches of
high energy photons (i.e, the lags of high-quality high energy light curves) to constrain the LIV.
Furthermore, since the emission mechanism of GRBs is still poorly understood, it is difficult to distinguish
an intrinsic time delay at the source from a delay induced by propagation in vacuum to the observer.
That is, the method of the flght-time difference used for testing LIV is hindered by our ignorance
concerning the intrinsic time delay in different energy bands
(see, e.g., \citealt{2006APh....25..402E,2009CQGra..26l5007B}).

The first attempt to disentangle the intrinsic time delay problem was presented in \cite{2006APh....25..402E}.
They proposed to work on statistical samples of GRBs at a range of different redshifts, and formulated
the problem in terms of a linear regression analysis where the slope corresponds to the QG
scale related to the LIV effect, and the intercept represents the possible intrinsic time delay. This analysis
has the advantage that it can extract the spectral time lags of broad light curves in different energy bands.
In this manner, a weak evidence for LIV was found under the assumption that all GRBs had the same intrinsic
time delay \citep{2006APh....25..402E}.
However, due to the fact that the durations of GRBs span about six orders of magnitude, it is not likely
that the high energy photons emitted from different GRBs (or from the same GRB) have the same
intrinsic time lag as compared with the emission time of the low energy photons \citep{2016ChPhC..40d5102C}.
As an improvement, \cite{2015APh....61..108Z} fitted the data of the energetic photons from GRBs on straight
lines with the same slope but with different intercepts (i.e., different intrinsic time lags). Unfortunately,
photons from different GRBs on the same line still mean that the intrinsic time lags between the high
energy photons and the onset low energy photons are approximately the same for these GRBs, which is not
always true and it could be a coincidence. \cite{2012APh....36...47C} estimated the intrinsic time lag between
emissions of high and low energy photons from GRBs by using the magnetic jet model. However, the magnetic
jet model relies on some particular theoretical parameters, and this leads to uncertainties on the LIV results.

Recently, \emph{Fermi} detected a peculiar burst GRB 160625B, which had three dramatically
different isolated sub-bursts \citep{2016GCN..19581...1B}, with unusually high photon statistics allowing
the use of amply populated energy bands. Here we calculate the spectral lags between the lowest energy band
(10--12 keV) and any other high energy band for the second sub-burst of GRB 160625B,
and find that the lag increases at $E\la8$ MeV, and then shows a steep decline in the energy range
$8\,{\rm MeV}\la  E\la20$ MeV. In other words, the behavior of the spectral lags of this GRB is quite different,
and a transition from \emph{positive} lags to \emph{negative} lags is, for the first time, discovered within a
burst.  If the LIV effect which happens at high energy scales is considered here, the observed time lag
($\Delta t_{\rm obs}$) between different energy bands of a GRB should consist of two terms
\begin{equation}
\Delta t_{\rm obs}=\Delta t_{\rm int} + \Delta t_{\rm LIV} \;,
\label{eq:tobs}
\end{equation}
where $\Delta t_{\rm int}$ represents the intrinsic emission time delay, and $\Delta t_{\rm LIV}$
denotes the time delay induced by the LIV effect. Instead of assuming an unknown constant
for $\Delta t_{\rm int}$, we argue that the intrinsic lag should be positively correlated with the energy,
i.e., the higher-energy photon arrives earlier than the lower-energy photon \citep{2016arXiv161007191S}.
Due to the LIV effect at high energy scales, a high-energy photon emitted (ideally) simultaneously  with
a low energy photon is to be observed later than that low-energy photon. Here we study the LIV effect of the
high energy photons in this direction. Put differently, $\Delta t_{\rm int}$ and $\Delta t_{\rm LIV}$ have
a different sign and, therefore, this positive correlation between the lag and the energy gradually
becomes an anti-correlation.

In this work, we develop a new method through which a reasonable formulation of the intrinsic time delay
can be derived, by fitting the energy dependence of the time lag. This allows us simultaneously
to obtain robust limits on the 1st order and 2nd order QG energy scale. We describe the data
analysis in Section~\ref{sec:obser}, and our methods and results are presented in Section~\ref{sec:LIV}.
Our conclusions are briefly summarized in Section~\ref{sec:summary}.

\section{The Observed Properties of Spectral Lags of GRB 160625B}
\label{sec:obser}

At $T_{0}=22:40:16.28$ UT on 2016 June 25, the \emph{Fermi} Gamma-Ray Burst Monitor (GBM; \citealt{2016GCN..19581...1B}) triggered
and located GRB 160625B for the first time. Then the \emph{Fermi} Large Area Telescope (LAT; \citealt{2016GCN..19580...1D})
detected a sharp increase in the rate of high-energy photons at 22:43:24.82 UT, resulting in an onboard
trigger on a bright pulse from the same GRB. At 22:51:16.03 UT, GBM triggered again on GRB 160625B \citep{2016GCN..19581...1B}.
The gamma-ray light curve of GRB 160625B consists of three dramatically different isolated sub-bursts with a total
duration of about $T_{90}=770$ s (15--350 keV; \citealt{2016arXiv161203089Z}). The first sub-burst
(precursor) that initially triggered the GBM is soft with a duration of about 0.84 s. The precursor is followed,
corresponding to the LAT trigger and starting at $\sim T_{0}+180$ s, by the main, extremely bright and spectrally-hard
episode with a duration of about 35.10 s.
After a long waiting time ($\sim 339$ s), the third sub-burst trigger GBM again which has
a duration of about 212.22 s.
Spectroscopic observations reveals absorption features consistent with Mg I, Mg II, Mn II and Fe II lines
at a redshift of $z=1.41$ \citep{2016GCN..19600...1X}.

\begin{figure}
\centerline{\includegraphics[angle=0,width=0.45\textwidth]{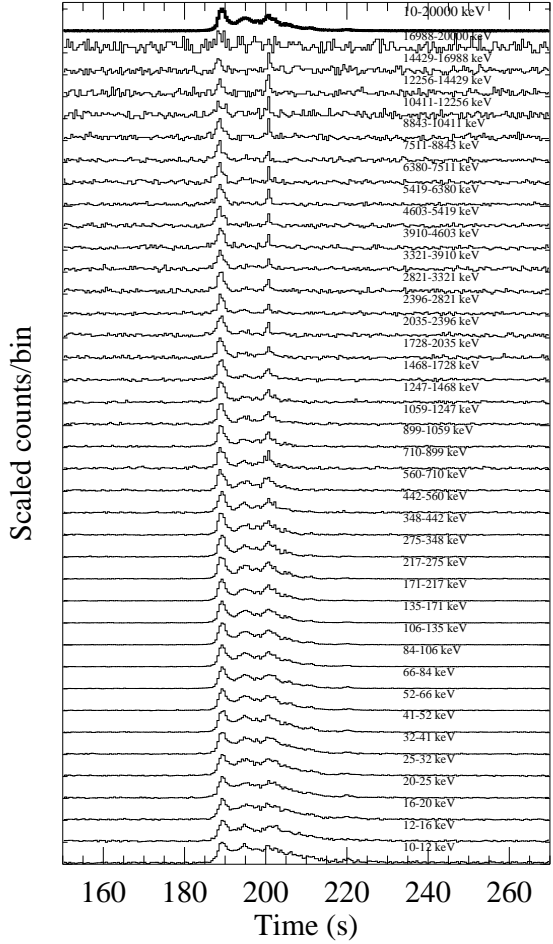}}
\vskip-0.1in
\caption{Energy-dependent light curves of the second sub-burst of GRB 160625B.
The full-range (10--20000 keV) light curve is shown on top (thick black line).}
\label{f1}
\end{figure}
\begin{figure}
\vskip-0.1in
\centerline{\includegraphics[angle=0,width=0.6\textwidth]{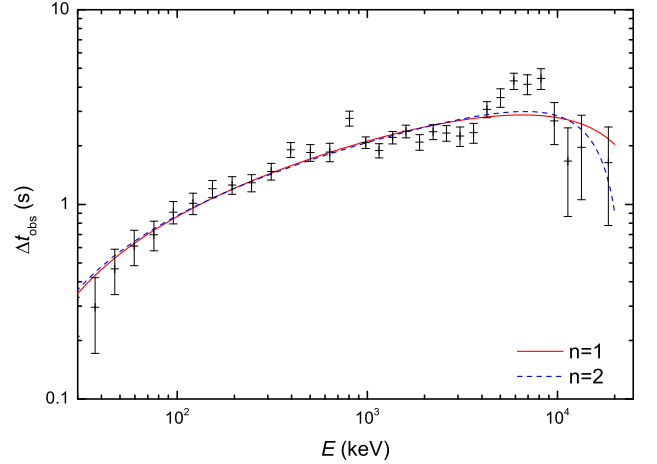}}
\vskip-0.1in
\caption{Energy dependence of the observed spectral lag $\Delta t_{\rm obs}$ (relative to the softest band),
and the best-fit theoretical curves:
(solid line) the linear ($n=1$) LIV model; (dashed line) the quadratic ($n=2$) LIV model.}
\label{f2}
\end{figure}

Since the second sub-burst of GRB 160625B is very bright, we can easily extract its light curves in
different energy bands (see Figure~\ref{f1}). In this analysis, we use the cross-correlation function (CCF) method to
calculate the lags between light curves of different energies for intervals 180.6--215.7 s of the burst.
The detailed CCF method is described in \cite{2012ApJ...748..132Z}. We look for spectral time lags
in the light curves recorded in the lowest energy band (10--12 keV) relative to any other GBM light curves
with higher energy bands, and find that the lag behavior is quite different. A transition from $positive$ lags
to $negative$ lags is first discovered at $E\sim8$ MeV (see Figure~\ref{f2}). The observed time lags we extract from
the energy-dependent light curves are listed in Table~\ref{table1}, together with their energy bands.


\begin{table}
\centering \caption{The time lags between the lowest energy band (10--12 keV)
and any other high energy bands for the second sub-burst of GRB 160625B}
\begin{tabular}{cc|cc}
\hline
\hline
 Energy & $\Delta t_{\rm obs}$ &  Energy & $\Delta t_{\rm obs}$ \\
 (keV) & (s) & (keV) & (s)  \\
\hline
12--16	&	$	-0.070	\pm	0.134	$	&	1059--1247	&	$	1.892	\pm	0.158	$	\\
16--20	&	$	-0.015	\pm	0.130	$	&	1247--1468	&	$	2.208	\pm	0.162	$	\\
20--25	&	$	0.081	\pm	0.125	$	&	1468--1728	&	$	2.375	\pm	0.179	$	\\
25--32	&	$	0.210	\pm	0.123	$	&	1728--2035	&	$	2.088	\pm	0.193	$	\\
32--41	&	$	0.296	\pm	0.124	$	&	2035--2396	&	$	2.361	\pm	0.208	$	\\
41--52	&	$	0.466	\pm	0.122	$	&	2396--2821	&	$	2.325	\pm	0.212	$	\\
52--66	&	$	0.611	\pm	0.127	$	&	2821--3321	&	$	2.242	\pm	0.255	$	\\
66--84	&	$	0.699	\pm	0.122	$	&	3321--3910	&	$	2.334	\pm	0.273	$	\\
84--106	&	$	0.913	\pm	0.120	$	&	3910--4603	&	$	3.080	\pm	0.290	$	\\
106--135	&	$	1.012	\pm	0.128	$	&	4603--5419	&	$	3.538	\pm	0.382	$	\\
135--171	&	$	1.204	\pm	0.121	$	&	5419--6380	&	$	4.306	\pm	0.409	$	\\
171--217	&	$	1.257	\pm	0.132	$	&	6380--7511	&	$	4.142	\pm	0.483	$	\\
217--275	&	$	1.290	\pm	0.131	$	&	7511--8843	&	$	4.435	\pm	0.542	$	\\
275--348	&	$	1.477	\pm	0.145	$	&	8843--10411	&	$	2.681	\pm	0.653	$	\\
348--442	&	$	1.908	\pm	0.168	$	&	10411--12256	&	$	1.670	\pm	0.803	$	\\
442--560	&	$	1.846	\pm	0.182	$	&	12256--14429	&	$	1.962	\pm	0.906	$	\\
560--710	&	$	1.856	\pm	0.202	$	&	14429--16988	&	$	-0.223	\pm	1.030	$	\\
710--899	&	$	2.765	\pm	0.243	$	&	16988--20000	&	$	1.637	\pm	0.858	$	\\
899--1059	&	$	2.079	\pm	0.143	$	&				&	$				$	\\
\hline
\end{tabular}
\label{table1}
\end{table}

\section{Constraints on the Violation of Lorentz Invariance}
\label{sec:LIV}
As mentioned above, the LIV-induced time lag $\Delta t_{\rm LIV}$ may be accompanied by an unknown
intrinsic energy-dependent time lag $\Delta t_{\rm int}$ caused by the unknown emission mechanism of GRBs.
Here we propose, for the first time, that the contributions of both the intrinsic time lag and the LIV effects
can result in a  lag behavior with a transition from $positive$ lags to $negative$ lags. Due to the fact that
the dominant fraction of GRB light curves show $positive$ lags (e.g., \citealt{2016arXiv161007191S}), we suggest
that there is a positive correlation between the intrinsic time lag and the energy. As the LIV effect takes the
lead at the high energy scales, this positive correlation trends in an opposite way.

\subsection{The intrinsic energy-dependent Time Lag}
In the observer frame, we assume the intrinsic positive time lag (between the lowest energy band and any other
high energy bands) increases with the energy $E$ in the form of an approximate power-law function:
\begin{equation}
\Delta t_{\rm int}(E)=\tau\left[\left(\frac{E}{\rm keV}\right)^{\alpha}-\left(\frac{E_{0}}{\rm keV}\right)^{\alpha}\right]\;{\rm s} \;,
\label{eq:tint}
\end{equation}
with $\tau>0$ and $\alpha>0$, where $E_{0}=11.34$ keV is the median value of the fixed lowest energy band (10--12 keV).
The coefficient $\tau$ and the index $\alpha$ are free parameters, which must be optimized simultaneously with
the QG energy scale $E_{\rm QG}$ (more on this below). We emphasize that a $positive$ lag corresponds to
an earlier arrival time for the higher energy emission in this study.

\subsection{The time delay induced by the LIV effect}
In QG scenarios, the LIV induced modifications to the photon dispersion relation can be expressed by
the leading term of the Taylor expansion as
\begin{equation}
E^{2}\simeq p^{2}c^{2}\left[1-s_{\pm}\left(\frac{pc}{E_{{\rm QG},n}}\right)^{n}\right]\;,
\label{eq:dispersion}
\end{equation}
which corresponds to a photon propagation speed
\begin{equation}
v=\frac{\partial E}{\partial p}\approx c\left[1-s_{\pm}\frac{n+1}{2}\left(\frac{E}{E_{{\rm QG},n}}\right)^{n}\right]\;,
\end{equation}
where $E_{\rm QG}$ denotes the QG energy scale, the $n$-th order expansion of the leading term
corresponds to the linear ($n=1$) or quadratic ($n=2$ order), and $s_{\pm}=\pm1$ is the ``sign" of the LIV  effect
($s_{\pm}=+1$ or $s_{\pm}=-1$ stands for a decrease or an increase in photon velocity with an increasing photon
energy). In the case of $s_{\pm}=+1$, photons with higher energies would travel slower than those with lower
energies in vacuum. This predicts a $negative$ spectral lag due to LIV, so we consider the $s_{\pm}=+1$ case
in the following.

Because of the energy dependence of the photon speed, two photons with different energies (denoted by $E$
and $E_{0}$, where $E>E_{0}$) emitted simultaneously from the same source would arrive on Earth
at different times. Taking into account the cosmological expansion, the LIV induced time delay is given by
\citep{2008JCAP...01..031J,2015APh....61..108Z}
\begin{equation}
\begin{aligned}
\Delta t_{\rm LIV}&=t_{\rm l}-t_{\rm h}\\
                  &=-\frac{1+n}{2H_{0}}\frac{E^{n}-E_{0}^{n}}{E_{{\rm QG}, n}^{n}}
\int_{0}^{z}\frac{(1+z')^{n}dz'}{\sqrt{\Omega_{\rm m}(1+z')^{3}+\Omega_{\Lambda}}}\;,
\label{eq:tLIV}
\end{aligned}
\end{equation}
where $t_{\rm l}$ and $t_{\rm h}$ are the arrival times of the low energy photon and the high energy photon, respectively.
Here we use the cosmological parameters determined by the $\it Planck$ observations \citep{2014A&A...571A..16P}:
$H_{0}=67.3$ km $\rm s^{-1}$ $\rm Mpc^{-1}$, $\Omega_{\rm m}=0.315$, and $\Omega_{\Lambda}=1-\Omega_{\rm m}$.

\subsection{Results}
With the 37 lag--energy measurements (see Table~\ref{table1} and Figure~\ref{f2}), from Equations~(\ref{eq:tobs},
\ref{eq:tint}, and \ref{eq:tLIV}), we perform a global fitting to
determine the free parameters ($\tau$, $\alpha$, and $E_{\rm QG}$) simultaneously using the Monte Carlo (MC) approach.
A fitting engine ($McFit$) has been developed, which employs a Bayesian MC fitting technique to realistically fit
free parameters that are constrained by the observed data even when other parameters are unconstrained. With the help
of this technique, the best-fit parameters and their uncertainties can be reliably determined by the converged
MC chains.

\begin{figure}
\centerline{\includegraphics[angle=0,width=0.5\textwidth]{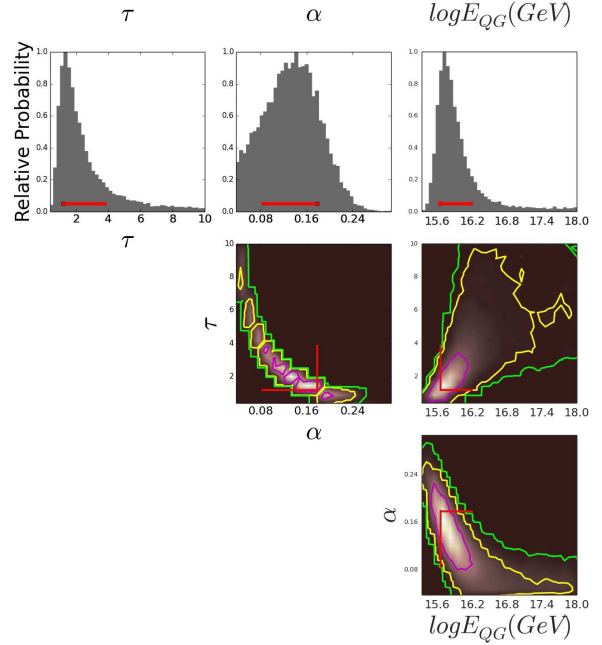}}
\vskip-0.1in
\caption{Parameter constraints of the linear ($n=1$) LIV case fitting for the lag--energy data.
Histograms and contours display the likelihood map of the parameter-constraint outputs by our $McFit$ package.
Red crosses indicate the best-fit values and their $1\sigma$ error bars.}
\label{f3}
\end{figure}

We first fit the observed lag--energy data with the linear LIV case (i.e., $n=1$). The resulting constraints on
$\tau$, $\alpha$, and $E_{\rm QG,1}$ are shown in Figure~\ref{f3}. These contours show that at the $1\sigma$ level,
the best-fit parameter values are $\tau=1.20^{+2.71}_{-0.04}\;{\rm s}$, $\alpha=0.18^{+0.01}_{-0.10}$,
and $\log (E_{\rm QG,1}/ \rm GeV)=15.66^{+0.55}_{-0.01}$, with a $\chi^{2}_{\rm dof}=81.22/34=2.39$.

\begin{figure}
\centerline{\includegraphics[angle=0,width=0.5\textwidth]{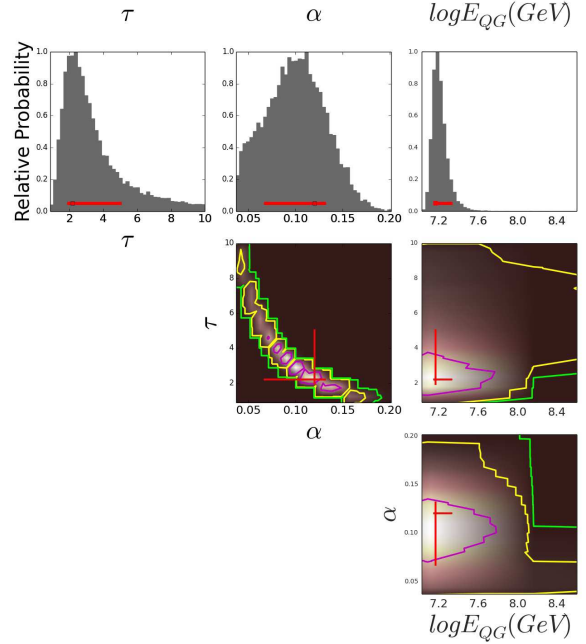}}
\vskip-0.1in
\caption{Same as Figure~\ref{f3}, except now for the quadratic (n=2) LIV case.}
\label{f4}
\end{figure}

Next, we consider the quadratic LIV case (i.e., $n=2$) to fit the observed lag--energy data.
The parameter constraints are displayed in Figure~\ref{f4}. We see here that the best-fit corresponds to
$\tau=2.18^{+2.90}_{-0.31}\;{\rm s}$, $\alpha=0.12^{+0.01}_{-0.05}$,
and $\log (E_{\rm QG,2}/ \rm GeV)=7.17^{+0.17}_{-0.02}$. With $37-3=34$ degrees of freedom,
we have $\chi^{2}_{\rm dof}=76.59/34=2.25$.

The best-fitting theoretical curves for the linear LIV case (solid line; with $\tau=1.20$ s, $\alpha=0.18$,
and $\log (E_{\rm QG,1}/ \rm GeV)=15.66$) and for the quadratic LIV case (dashed line; with $\tau=2.18$ s,
$\alpha=0.12$, and $\log (E_{\rm QG,2}/ \rm GeV)=7.17$) are shown in Figure~\ref{f2}. This reveals that
both cases are adequate to represent the data.

With the best-fit values of $\log E_{\rm QG,1}$ and $\log E_{\rm QG,2}$ as well as their $1\sigma$ error bars,
the $1\sigma$ confidence-level lower limit on LIV is
\begin{equation}
E_{\rm QG,1}\geq0.5\times10^{16}\; \rm GeV
\end{equation}
for the linear ($n=1$) LIV case, and
\begin{equation}
E_{\rm QG,2}\geq1.4\times10^{7}\; \rm GeV
\end{equation}
for the quadratic ($n=2$) LIV case.

\subsection{Other possible LIV tests}
There are some observations (e.g, the ultrahigh energy cosmic rays and the TeV photons) with energies above
an expected threshold (pion production or pair creation), which can be considered as ``threshold anomalies".
It has been proposed that LIV could be the origin of the threshold anomalies \citep{2001PhRvD..64c6005A}.
That is, the LIV scenario could be tested not only by the study of spectral lags but also by analysis of data on
such threshold anomalies. The limits on LIV from the threshold anomalies are at a level well beyond the
sensitivity of the analysis reported in this paper (see \citealt{2005LRR.....8....5M,2013LRR....16....5A}, and
references therein).  On the other hand, it is well known that threshold anomalies do not apply to the DSR scenario
(see \citealt{2005LRR.....8....5M,2013LRR....16....5A}).  These results on threshold anomalies would imply that the
analysis of spectral lags reported in this paper would not carry much weight on the debate concerning LIV (since
more stringent limits may be established via the threshold anomalies), being instead a more valuable
contribution to the debate on the DSR scenario.

\section{Summary and Discussion}
\label{sec:summary}

The observed spectral lags of GRBs have been widely used to constrain the energy scales of LIV. The key issue
in the idea of searching for spectral lags, however, is to distinguish the possible time delay induced by
the LIV effect from any source-intrinsic time lag in the emission of photons at different energies.
In order to overcome the intrinsic time lag problem, \cite{2006APh....25..402E} proposed a data fitting procedure
to test the LIV effect, and an unknown constant was assumed to be the intrinsic time lag in the linear fitting
function, and furthermore assuming that all GRBs have the same intrinsic time lag.

Here, instead of assuming an unknown constant for the intrinsic time lag, we argue that the intrinsic lag has a
positive dependence on the photon energy. On the other hand, the LIV effects  which are expected at high energy
scales would make high energy photons travel in vacuum slower than low energy photons, so we suggest that the
positive correlation between the lag and the energy will gradually become an anti-correlation at the high energy
scales. In this work, we successfully fit the evolving behavior of the spectral lags of GRB 160625B (i.e., the
existence of a transition from $positive$ lags to $negative$ lags), by considering the contributions of both
the intrinsic time lag and the lag by the LIV effect. This is the first time, to our knowledge, that it
has been possible to give both a reasonable formulation of the intrinsic energy-dependent time lags and
robust limits on LIV through direct fitting of the spectral lag data of a GRB.

Our limit on the linear LIV case ($E_{\rm QG,1}\geq0.5\times10^{16}$ GeV) obtained here from the spectral lags
is comparable to the limit found from \cite{2006APh....25..402E} with an unknown constant for the intrinsic time
lag, being less than three orders of magnitude below the Planck energy scale. Our limit on the quadratic LIV case
($E_{\rm QG,2}\geq1.4\times10^{7}$ GeV) is four orders of magnitude below the current best limit from the single
GeV photon of GRB 090510 \citep{2009Natur.462..331A,2013PhRvD..87l2001V}. While the spectral lags of GRB 160625B
do not currently have the best sensitivity to LIV constraints, there is nonetheless merit to the result.
Firstly, because the true spectral time lags of broad light curves in different energy multi-photon bands are
used to obtain reliable constraints on LIV, rather than the rough time lags obtained from a single GeV photon.
Secondly, because the analysis of the intrinsic time lag performed here is important for studying the flight time
differences from the astronomical sources to test the LIV effect, since it impacts the reliability of the
resulting constraints on LIV. Compared with previous works, the problems associated with the intrinsic time lags
can be obviously handled better with our new method. Furthermore, it is reasonable to expect that GRB 160625B is
not the only burst where a transition from $positive$ to $negative$ lags can be determined, and the method
presented here can be used for any burst with similar lag features. More stringent constraints on LIV can be
expected as our method is applied to larger numbers of GRBs with higher temporal resolutions and more high
energy photons.

\acknowledgments
We thank the anonymous referee for constructive suggestions.
This work is supported by the National Basic Research Program (``973" Program)
of China (Grant No 2014CB845800), the National Natural Science Foundation of China
(Grant Nos. 11322328, 11433009, 11673068, 11603076, and 11103083), the Youth Innovation Promotion
Association, the Key Research Program of Frontier Sciences (QYZDB-SSW-SYS005),
the Strategic Priority Research Program ``Multi-waveband gravitational wave Universe"
(Grant No. XDB23000000) of the Chinese Academy of Sciences, the Natural Science Foundation
of Jiangsu Province (Grant No. BK20161096), and the Guangxi Key Laboratory for Relativistic Astrophysics.
BBZ acknowledges support from the Spanish Ministry Projects AYA 2012-39727-C03-01 and AYA2015-71718-R.
Part of this work used BBZ's personal IDL code library ZBBIDL and personal Python library ZBBPY.
The computation resources used in this work are owned by Scientist Support LLC.
L.S. acknowledges support from the Joint NSFC-ISF Research Program (No. 11361140349),
jointly funded by the National Natural Science Foundation of China and the Israel Science Foundation.
P.M. acknowledges NASA NNX 13AH50G.


\end{document}